\documentclass[11pt]{article}
\usepackage{a4}
\usepackage{times}
\usepackage{amssymb}

\def\beq{\begin{equation}}
\def\bea{\begin{eqnarray}}
\def\eeq{\end{equation}}
\def\eea{\end{eqnarray}}

\def\Z{$\mathbb{Z}$}
\def\SM{Standard Model}

\begin{document}
\input{epsf.sty}
\title{\bf  Almost the supersymmetric Standard Model from intersecting D6-branes on the $\mathbb{Z} _6'$ orientifold }

\maketitle
\vglue 0.35cm
\begin{center}
\author{\bf David Bailin \footnote
{D.Bailin@sussex.ac.uk} \&  Alex Love \\}
\vglue 0.2cm
	{\it  Department of Physics \& Astronomy, University of Sussex\\}
{\it Brighton BN1 9QH, U.K. \\}
\baselineskip=12pt
\end{center}
\vglue 2.5cm
\begin{abstract}
Intersecting stacks of supersymmetric fractional branes on the \Z$_6'$ orientifold 
may be used to construct the supersymmetric Standard Model.
   If $a,b$  are the   stacks 
  that generate the  $SU(3)_{\rm colour}$ and $SU(2)_L$ gauge particles,  
 then,  in order to obtain {\em just} the chiral spectrum of the (supersymmetric)
  Standard Model (with non-zero Yukawa couplings to the Higgs mutiplets),
   it is necessary that  the number of intersections $a \cap b$ of the stacks $a$ and $b$, and 
  the number of intersections $a \cap b'$ of $a$ with the orientifold image $b'$ of $b$
   satisfy $(a \cap b,a \cap b')=(2,1)$ or $(1,2)$. 
It is also necessary that there is no matter in symmetric representations of the gauge group,
 and not too much matter in antisymmetric representations, on either stack. 
 Fractional branes having  all of these properties may be constructed  on the \Z$_6'$ orientifold. 
  We construct a (four-stack)
  model with two  further stacks, each 
 with just a single brane, which has precisely the matter spectrum of the  supersymmetric Standard Model, 
 including a single pair of Higgs doublets. However, 
 the gauge group is $SU(3)_{\rm colour} \times SU(2)_L \times U(1)_Y \times U(1)_H$. 
 Only the Higgs doublets are charged with respect to $U(1)_H$.
 \end{abstract}
\newpage
An attractive, bottom-up approach to constructing the Standard Model  is to  use intersecting  D6-branes \cite{Lust:2004ks}. 
In these models one starts with 
 two stacks, $a$ and $b$ with $N_a=3$ 
and $N_b=2$, of   D6-branes wrapping the three large spatial 
dimensions plus  3-cycles of the six-dimensional  internal space (typically a torus $T^6$ 
or a Calabi-Yau 3-fold) on which the theory is compactified.
 These generate  the gauge group $U(3) \times U(2) \supset SU(3) _c \times SU(2)_L$, 
 and  the non-abelian component of the standard model gauge group
is immediately assured.
   Further, (four-dimensional) fermions in bifundamental representations 
$({\bf N} _a, \overline{\bf N}_b)= ({\bf 3}, \overline{\bf 2})$ 
of the gauge group can arise at the multiple intersections of the two stacks. 
These are precisely the representations needed for the quark doublets $Q_L$ of the Standard Model, 
and indeed an 
attractive model having just the spectrum of the Standard Model has been constructed \cite{Ibanez:2001nd}. The D6-branes wrap 3-cycles 
of an orientifold $T^6/\Omega$, where $\Omega$ is the world-sheet parity operator. The advantage and, indeed, the necessity of using 
an orientifold stems from the fact that for every stack $a,b, ...$ there is an orientifold image $a',b', ...$ . 
At intersections of $a$ and $b$ there are chiral fermions 
in the $({\bf 3}, \overline{\bf 2})$ representation of $U(3) \times U(2)$, where the ${\bf 3}$ has charge $Q_a=+1$ with respect to the 
$U(1)_a$ in $U(3)=SU(3)_{\rm colour} \times U(1)_a$, and the $\overline{\bf 2}$ has charge $Q_b=-1$ with respect to the 
$U(1)_b$ in $U(2)=SU(2)_L \times U(1)_b$.  However, at intersections of $a$ and $b'$ there are chiral fermions 
in the $({\bf 3},{\bf 2})$ representation, where  the ${\bf 2}$ has $U(1)_b$ charge $Q_b=+1$. 
In the model of \cite{Ibanez:2001nd}, the number of intersections $a \cap b$ of the stack $a$ with $b$ is 2, 
and the number of intersections $a \cap b'$ of the stack $a$ with $b'$ is 1. Thus, as required for the \SM , there are 3 quark doublets. (They have the same weak hypercharge $Y$ provided that $Q_b$ does not contribute to $Y$.)
 These have  net $U(1)_a$ charge $Q_a=6$, and net $U(1)_b$ charge $Q_b=-3$. Tadpole cancellation requires that overall both charges,
sum to zero, so further fermions are essential, and indeed required by the \SM. 6 quark-singlet states $u^c_L$ and $d^c_L$ 
belonging to the $({\bf 1}, \overline{\bf 3})$ representation of  $U(1) \times U(3)$, having 
a total of $Q_a=-6$ are sufficient to ensure overall cancellation of $Q_a$, and these arise from the intersections of $a$ with other 
stacks $c,d,...$ having just a single D6-brane. Similarly, 3 lepton doublets $L$, belonging to the $({\bf 2}, \overline{\bf 1})$
 representation of  $U(2) \times U(1)$, having 
a total $U(1)_b$ charge of $Q_b=3$, are sufficient to ensure overall cancellation of $Q_b$, 
and these arise from the intersections of $b$ with other 
stacks having just a single D6-brane. In contrast, had we not used an orientifold, the requirement of 3 quark doublets would 
necessitate having the number of intersections $a \cap b=3$. This makes no difference to the charge $Q_a=6$ carried by the quark doublets,   
but instead the $U(1)_b$ charge carried by the quark doublets is $Q_b=-9$, which cannot be cancelled by just 3 lepton doublets $L$. 
Consequently, additional vector-like fermions are unavoidable unless the orientifold projection is available.
This is why the orientifold is essential if we are to get just the matter content of the \SM \ or of the MSSM. 

 Actually, an orientifold  can allow essentially the standard-model spectrum without vector-like matter even 
when $a \cap b=3$ and $a \cap b'=0$ \cite{Blumenhagen:2001te}. 
This is because in orientifold models it is also possible to get chiral matter in the symmetric and/or antisymmetric representation 
of the relevant gauge group from open strings stretched between a stack and its orientifold image. Both representations have charge $Q=2$ 
with respect to the relevant $U(1)$. The antisymmetric (singlet) representation of $U(2)$ can 
describe a neutrino singlet state $\nu ^c_L$ (since $Q_b$ does not contribute to $Y$),
 and 3 copies contribute $Q_b=6$ units of $U(1)_b$ charge. If there are also 3 lepton doublets $L$ belonging to the bifundamental representation 
$({\bf 2}, \overline{\bf 1})$
 representation of  $U(2) \times U(1)$, each contributing $Q_b=1$ as above, then the total contribution is $Q_b=9$ which {\bf can} 
 be cancelled by 3 quark doublets $Q_L$ in the $({\bf 3}, \overline{\bf 2})$ representation of $U(3) \times U(2)$. Thus,  
 orientifold models can allow
 the standard-model spectrum plus 3 neutrino singlet states even when $(a \cap b ,a \cap b')=(3,0)$.

Non-supersymmetric intersecting-brane models lead to 
 flavour-changing neutral-current  (FCNC) processes 
that can only be suppressed to levels consistent with the current bounds by making the 
 string scale  rather high, of order $10^4$ TeV, which in turn leads to fine-tuning problems  \cite{Abel:2003yh}.
Further, in  non-supersymmetric theories, such as these, the cancellation of Ramond-Ramond (RR) tadpoles does not ensure 
Neveu Schwarz-Neveu Schwarz (NSNS) tadpole cancellation. 
 Thus a particular consequence of the non-cancellation is that the complex structure moduli are unstable \cite{Blumenhagen:2001mb}. 
 One way to stabilise these moduli 
  is for the 
D-branes to wrap an orbifold $T^6/P$, where $P$ is a ``point group'' acting on $T^6$, rather than a torus $T^6$. 
The FCNC problem can be solved and the complex structure moduli stabilised when the theory is supersymmetric. 
First, a supersymmetric theory is not obliged to have the low string scale that led to problematic FCNCs  induced by string instantons. 
Second, 
 in a supersymmetric theory, RR tadpole cancellation ensures cancellation 
of the NSNS tadpoles \cite{Cvetic:2001tj,Cvetic:2001nr}.
An orientifold is then constructed by quotienting the orbifold with the world-sheet parity operator $\Omega$.

In this paper we shall be concerned with the  orientifold having  point group $P=$\Z$_6'$. We showed in 
a previous paper \cite{Bailin:2006zf} that this {\em does} have (fractional) supersymmetric D6-branes $a$
 and $b$ with intersection numbers 
$(a \cap b, a \cap b')=(1,2)$ or $(2,1)$, which in principle
 might be used to construct the supersymmetric \SM \ having just the requisite
 standard-model matter content. 

The torus $T^6$ factorises into three 2-tori $T^2_1 \times T^2_2 \times T^2_3$, with 
$T^2_k \ (k=1,2,3)$  parametrised by the complex coordinate  $z_k$. 
 The action of the generator $\theta$ of the point group  $\mathbb{Z} ' _6$ on the   
coordinates $z_k $ is given by
\beq
\theta z_k = e^{2\pi i v_k} z_k
\eeq
where
\beq
 (v_1,v_2,v_3)= \frac{1}{6} (1,2,-3) 
 \label{z61vk}
\eeq
when $P=$\Z$_6'$. 
Since the point group action must be an automorphism of the lattice, we take $T^2_{1,2}$ to be $SU(3)$ lattices, 
so that their complex structure moduli  $U_{1,2}$ are given by $U_1= e^{i \pi/3}=U_2$. In contrast, the lattice for
 $T^2_3$, and hence its complex structure $U_3$, is  arbitrary. 
The anti-holomprphic embedding $\mathcal{R}$ of the world-sheet parity operator $\Omega$
 acts as complex conjugation on the coordinates $z_k$
 \beq
 \mathcal{R} z_k= \overline{z}_k \quad (k=1,2,3)
 \eeq
 Requiring that this too is an automorphism of the lattice 
 constrains the orientation of each torus $T^2_k$ relative to the Re $z_k$ axis.  Each torus must be
  in one of two configurations, denoted {\bf A} and {\bf B}, defined in reference \cite{Bailin:2006zf}. 
This fixes the real part  of the complex structure Re $U_3=0,1/2$ respectively,
 but the imaginary part Im $U_3$ remains {\it a priori} arbitrary.
  In this paper, we shall only be concerned with the 
  {\bf ABA} lattice.

The fractional branes $\kappa$ with which we are concerned have the general form
\beq
\kappa = \frac{1}{2} \left( \Pi _{\kappa}^ {\rm bulk} + \Pi _{\kappa}^ {\rm ex} \right)
\eeq
where 
\beq
\Pi _{\kappa}^ {\rm bulk}= \sum_{p=1,3,4,6} A^{\kappa}_p \rho _p
\eeq
is an (untwisted) invariant 3-cycle, and
\beq
\Pi _{\kappa}^ {\rm ex}= \sum_{j=1,4,5,6} (\alpha^{\kappa}_j \epsilon _j + \tilde{\alpha}^{\kappa}_j \tilde{\epsilon} _j )
\eeq
is an exceptional 3-cycle associated with the $\theta ^3$-twisted sector. It consists of a collapsed 2-cycle at a 
$\theta^3$ fixed point in $T^2_1 \times T^2_3$ times a 1-cycle in (the $\theta^3$-invariant plane) $T^2_2$. 
The 4 basis invariant 3-cycles $\rho _p, \ (p=1,3,4,6)$ and the 8 basis exceptional cycles $\epsilon _j$ and 
$\tilde{\epsilon}_j, (j=1,4,5,6)$ are defined in reference \cite{Bailin:2006zf}. Their non-zero intersection numbers are
\bea
\rho _1 \cap \rho _4=-4, &\quad & \rho _1 \cap \rho _6=2 \\
\rho _3 \cap \rho _4=2, &\quad & \rho _3 \cap \rho _6=-4 
\eea
and 
\beq
\epsilon _j \cap \tilde{\epsilon}_k= -2 \delta _{jk}
\eeq

The wrapping numbers $(n^a_k,m^a_k)$ of the basis 1-cycles $(\pi _{2k-1}, \pi _{2k})$ of $T^2_k$ for the $U(3)$ stack $a$ are given by
\beq
(n^a_1,m^a_1;n^a_2,m^a_2,n^a_3,m,^a_3)=(1,-2;-1,0;1,-2)
\eeq
Using the formulae given in \cite{Bailin:2006zf}, 
\bea 
A_1= (n_1n_2+n_1m_2+ m_1n_2)n_3  \label{A1}\\
A_3= (m_1m_2+n_1m_2+ m_1n_2)n_3 \\
A_4= (n_1n_2+n_1m_2+ m_1n_2)m_3 \\
A_6= (m_1m_2+n_1m_2+ m_1n_2)m_3 \label{A6}
\eea 
we can compute the bulk coefficients $A^a_p$ for $\Pi _a^{\rm bulk}$. Then 
the fractional brane $a$ has
\bea
\Pi _a^{\rm bulk}&=& \rho _1+2\rho_3-2\rho _4-4 \rho_6  \label{abulk}\\
\Pi _a^{\rm ex}&=&(-1)^{\tau ^a_0}\left(2[\epsilon _1 +(-1)^{\tau ^a _2}\epsilon _4]+
[\tilde{\epsilon} _1 +(-1)^{\tau ^a _2}\tilde{\epsilon} _4] \right) \label{aex}
\eea
On the {\bf ABA} lattice, the orientifold O6-plane is
\beq
\Pi _{\rm O6}=2 \rho _1+\rho _3-3 \rho _6 \label{pio6}
\eeq
and  the orientifold images of $\Pi _a^{\rm bulk}$ and $\Pi _a^{\rm ex}$ are
\bea
{\Pi _a^{\rm bulk}}'&=& \rho _1-\rho_3+2\rho _4-2 \rho_6 \\
{\Pi _a^{\rm ex}}'&=&=(-1)^{\tau ^a_0}\left(-[\epsilon _1 +(-1)^{\tau ^a _2}\epsilon _4]+
[\tilde{\epsilon} _1 +(-1)^{\tau ^a _2}\tilde{\epsilon} _4] \right) 
\eea
Then
\bea
a \cap \Pi _{\rm O6}&=&\frac{1}{2}\Pi _a^{\rm bulk} \cap \Pi _{\rm O6}=3  \label{pio6a} \\
&=&a \cap a'
\eea
from which it follows, as required, that there are no symmetric representations ${\bf S}_a={\bf 6}$ on the stack $a$. 
On this lattice, supersymmetry requires that the bulk coefficients $A^a_p$ satisfy 
\bea
\sqrt{3}A^a_1+(A^a_4-2A^a_6) \ {\rm Im} \ U_3>0 \\
-A^a_1+2A^a_3+A^a_4 \sqrt{3}  \ {\rm Im} \ U_3=0
\eea
where $U_3$ is the complex structure on $T^2_3$. Using the bulk coefficients for $a$ given in (\ref{abulk}), it follows that
\beq
{\rm Im} \ U_3=\frac{\sqrt{3}}{2}
\eeq
Likewise, 
the wrapping numbers $(n^b_k,m^b_k)$  for the $U(2)$ stack $b$ are given by
\beq
(n^b_1,m^b_1;n^b_2,m^b_2,n^b_3,m,^b_3)=(0,1;0,-1;0,1)
\eeq
and the fractional brane $b$ has
\bea
\Pi _b^{\rm bulk}&=& - \rho_6 ={\Pi _b^{\rm bulk}}' \label{bbulk}\\
\Pi _b^{\rm ex}&=&=(-1)^{\tau ^b_0+1}[\epsilon _1 +(-1)^{\tau ^b _2}\epsilon _5]=-{\Pi _b^{\rm ex}}' \label{bex}
\eea
It follows that $b$ too is supersymmetric and that
\beq
b \cap \Pi _{\rm O6}=0=b \cap b' \label{pio6b}
\eeq 
so,  again,  there are no symmetric representations ${\bf S}_b={\bf 3}$ on the stack $b$.
Then, as required,
\beq
(a \cap b, a \cap b')=(2,1) \quad  {\rm or} \quad (1,2)
\eeq
the former occurring when $\tau  ^a_0=\tau^b_0 \bmod2$ and the latter when $\tau ^a_0\neq\tau^b_0 \bmod2$.

The weak hypercharge $Y$ is a linear combination
\beq
Y= \sum _{\kappa}y_{\kappa}Q_{\kappa}
\eeq
 of the charges $Q_{\kappa}$ associated with the $U(1)_{\kappa}$ group for 
each stack $\kappa$.
As explained above, the intersections $a \cap b$ give chiral supermultiplets in the $({\bf 3}, \overline{\bf 2})$ representation of 
$U(3) \times U(2)$ and $a \cap b'$ give chiral supermultiplets in the $({\bf 3}, {\bf 2})$ representation. Since both of these 
give quark doublets $Q_L$ having $Y= 1/6$, we infer that
\bea
y_a &=& \frac{1}{6} \\
y_b&=&0
\eea
(As noted earlier, the stack $Q_b$ makes no contribution to $Y$.)
Using (\ref{pio6a})  we find that the number of antisymmetric representations on the stack $a$ is
\beq
 \#({\bf A}_a)=\frac{1}{2}(a \cap a'+a \cap \Pi _{\rm O6})=3
 \eeq
 Since the antisymmetric part  ${\bf A}_a=({\bf 3} \times {\bf 3})_{\rm antisymm}=\overline{\bf 3}$ for the group $SU(3)$,
  and since ${\bf 3}$ has $Q_a=1$, it follows that
 there are 3 $(\overline{\bf 3},{\bf 1})$ representations of $SU(3)_{\rm colour} \times SU(2)_L$ having $Y=1/3$. Thus, there are 3 $d^c_L$ quark-singlet states on the stack $a$. 
 Similarly, from (\ref{pio6b}), it follows that
\beq
 \#({\bf A}_b)=0
 \eeq
 so that there are no lepton-singlet $\nu^c_L$ states on the stack $b$. To complete the standard-model spectrum, it is therefore necessary to add further stacks, 
 $c,d,e,...$ each having just a single (fractional) brane $N_{c,d,e...}=1$.  In principle, it might be possible to obtain the full spectrum
  with the addition of just one further stack $c$ with $y_c= 1/2$. 
 The remaining 3 quark-singlet states $u^c_L$ having  weak hypercharge $Y=-2/3$ might arise at  the intersections of $a$ and $c$, the  3 lepton and  2 Higgs doublets with $Y= \pm 1/2$ from intersections of $b$ with $c$, with   the 3 charged lepton-singlet states $\ell ^c_L$ having $Y=1$  possibly arising as symmetric representations on $c$. 
 However, we have not so far been able to find such an example. The example that we shall present has two further $U(1)$ stacks. 
 
 The wrapping numbers   $(n^c_k,m^c_k)$ for the first of these stacks are
\beq
(n^c_1,m^c_1;n^c_2,m^c_2,n^c_3,m,^c_3)=(1,0;1,0;3,2)
\eeq
and the fractional brane $c$ has
\bea
\Pi _c^{\rm bulk}&=& 3\rho_1+2 \rho_4\\
\Pi _c^{\rm ex}&=&(-1)^{\tau ^c_0+1}\left(2[\epsilon _1 +(-1)^{\tau ^c _2}\epsilon _4]+
[\tilde{\epsilon} _1 +(-1)^{\tau ^c _2}\tilde{\epsilon} _4] \right)
\eea
Then $c$ is supersymmetric and 
\bea
{\Pi _c^{\rm bulk}}'&=&3 \rho _1+3\rho_3-2\rho _4-2 \rho_6 \\
{\Pi _c^{\rm ex}}'&=&=(-1)^{\tau ^c_0+1}\left(-[\epsilon _1 +(-1)^{\tau ^c _2}\epsilon _4]+
[\tilde{\epsilon} _1 +(-1)^{\tau ^c _2}\tilde{\epsilon} _4] \right)
\eea
Taking 
\bea
\tau ^a_0 + \tau ^c_0= 0 \bmod2 \label{tac0} \\
\tau ^a_2 + \tau ^c_2= 0 \bmod2  \label{tac2}
\eea
and using (\ref{abulk}) and (\ref{aex}) then gives
\beq
(a \cap c, a \cap c')=(0,-3)
\eeq
Hence, if
\beq
y_c= \frac{1}{2}
\eeq
we get just the required 3 quark-singlet states $u^c_L$ with weak hypercharge $Y= -2/3$. 
Further, the number $\#({\bf S}_c)$  of symmetric representations on the stack $c$ is
\bea
 \#({\bf S}_c)&=& \frac{1}{2} (c \cap c' - c \cap \Pi _{\rm O6}) \\
 &=&3
 \eea
so that we get just the required 3 charged lepton-singlet states $\ell ^c_L$  having weak hypercharge $Y= 1$. 
Similarly, using (\ref{bbulk}) and (\ref{bex}), we find that
\beq
(b \cap c, b \cap c')=(2,-1) \quad {\rm or} \quad (1,-2)
\eeq
the former occurring when $\tau ^b_0 + \tau ^c_0=0 \bmod 2$ and the latter when it is $1 \bmod 2$. Either way, 
this gives 3 lepton (or Higgs) doublets each having weak hypercharge $Y=-1/2$. Evidently a further stack is required to 
generate a pair of doublets with opposite weak hypercharges  $Y=1/2$ and $Y=-1/2$.

The wrapping numbers   $(n^d_k,m^d_k)$ for the second of the $U(1)$ stacks are
\beq
(n^d_1,m^d_1;n^d_2,m^d_2,n^d_3,m,^d_3)=(1,0;1,1;1,0)
\eeq
and the fractional brane $d$ has
\bea
\Pi _d^{\rm bulk}&=& 2\rho_1+ \rho_3 = {\Pi _d^{\rm bulk}}'\\
\Pi _d^{\rm ex}&=&(-1)^{\tau ^d_0+1}\left([\epsilon _1 +(-1)^{\tau ^d _2}\epsilon _4]+
2[\tilde{\epsilon} _1 +(-1)^{\tau ^d _2}\tilde{\epsilon} _4] \right)={\Pi _d^{\rm ex}}'
\eea
Then $d$ is supersymmetric and  
\beq
d \cap d'=0=d \cap \Pi_{\rm O6}
\eeq
so that there are no (lepton-singlet) states arising as the symmetric representation ${\bf S}_d$ on the stack $d$. Also, 
taking 
\bea
\tau ^a_0 + \tau ^d_0= 0 \bmod2 \label{tad0} \\
\tau ^a_2 + \tau ^d_2= 0 \bmod2  \label{tad2}
\eea
and using (\ref{abulk}) and (\ref{aex}) then gives
\beq
(a \cap d, a \cap d')=(0,0)
\eeq
Thus there are no (unwanted) quark-singlet states arising at the intersections of $a$ with $d$ and $d'$. 
Using (\ref{bbulk}) and (\ref{bex}), we find that 
\beq
(b \cap d, b \cap d')=(1,1) \quad {\rm or} \quad (-1,-1) \label{Higgs}
\eeq
the former occurring when $\tau ^b_0 +\tau ^d_0= 0 \bmod 2$ and the latter when it is $1 \bmod 2$. Either way, a  
pair of  (Higgs) doublets, having the required opposite weak hypercharges  occurs, provided that 
\beq
y_d= \pm \frac{1}{2}
\eeq
Finally, using (\ref{tac0}),(\ref{tac2}),(\ref{tad0}) and (\ref{tad2}), we find that
\beq
(c \cap d, c \cap d')=(0,0)
\eeq
so  that there are no charged or neutral lepton singlets at the intersections of $c$ with $d$ or $d'$. 

In total, we have {\em just} the spectrum of the  supersymmetric  \ \SM, with a single pair of Higgs doublets, and no 
neutrino singlet states $\nu ^c_L$. For this to be a consistent string theory realisation of the \SM  \ it is necessary that 
there is overall cancellation of the RR tadpoles, and this in turn requires that the overall homology class of the 
D6-branes and O6-plane must vanish:
\beq
\sum _{\kappa} N_{\kappa}(\kappa + \kappa ')= 4 \Pi _{\rm O6}
\eeq
The  sum is  over all four stacks $\kappa = a,b,c,d$ and $\Pi _{\rm O6}$ is given in (\ref{pio6}). 
Note that the left-hand side has contributions from both bulk and exceptional D6-branes, whereas the right-hand side has only 
the former. Both bulk and exceptional parts are required to cancel separately, and 
it is easy to verify that this is the case. 

In the first instance the gauge group derived from these stacks is
\bea
G&=&U(3)_a \times U(2)_b \times U(1)_c \times U(1)_d \\
&=&SU(3)_{\rm colour} \times SU(2)_L \times U(1)_Y \times U(1)^3
\eea
The last three $U(1)$s are of course unwanted. The gauge boson of any anomalous $U(1)$ gauge group acquires a mass of order the 
string scale, O$(10^{17})$ GeV in a supersymmetric theory, and the $U(1)$ survives only as a global symmetry. This is 
what happens in the case of $U(1)_a$, which is the $U(1)$ associated with (3 times the) baryon number. However,  there remains the 
possibility that the  gauge boson of a non-anomalous $U(1)$ symmetry remains massless (on the string scale) and might therefore 
be observable in low-energy experiments. The  $U(1)$ gauge boson associated with a general linear combination 
of the $U(1)$ charges $Q_{\kappa}$
\beq
X = \sum _{\kappa} x_{\kappa} Q_{\kappa}
\eeq
whether anomalous or non-anomalous,
 does {\em not} acquire a mass via the Green-Schwarz mechanism provided that
\beq
\sum_{\kappa}x_{\kappa}N_{\kappa}(\kappa - \kappa ')=0
\eeq
In  the case under consideration, it is easy to verify that $U(1)_Y$ remains massless, as required, but so too does $U(1)_d$. 
Thus we have an unwanted further $U(1)$ factor in the gauge group. However, the only matter which has $Q_d \neq 0$ is the 
pair of doublets in (\ref{Higgs}) and it is attractive to identify these with the pair of Higgs doublets, $H$ and $\overline{H}$. 
Since, Higgs particles have not yet been observed, this scenario has not yet been falsified. 

In summary, we have found an attractive and economical realisation of the supersymmetric \SM \ using just four stacks of D6-branes 
on the {\bf ABA} lattice. It has the correct matter content, with no neutrino-singlet states. There is one additional $U(1)$ 
factor in the gauge group, which may only interact with Higgs doublets. 
Although the complex structure moduli are stabilised in this model, there of course remain unstabilised K\"ahler and 
dilaton  moduli. In principle, these may be stabilised using background fluxes, perhaps via the ``rigid corset'' proposed 
 in \cite{Camara:2005dc}. In any case, fluxes are presumably needed to break the $\mathcal{N}=1$ supersymmetry of the 
spectrum, as well as the $\mathcal{N}=2$ supersymmetry of the gauge supermultiplets. These matters will be 
discussed elsewhere. Likewise, 
the results of a systematic investigation of 
other possible lattices for the $\mathbb{Z} _6'$ orientifold will be presented in a separate paper \cite{BLZ612}. 


\begin{thebibliography}{99999}

\bibitem{Lust:2004ks}For a review, see 
  D.~L\"ust,
  Intersecting brane worlds: A path to the standard model?,
  Class.\ Quant.\ Grav.\  {\bf 21} (2004) S1399
  [hep-th/0401156].




  
\bibitem{Ibanez:2001nd}
  L.~E.~Ib\'a\~nez, F.~Marchesano and R.~Rabad\'an,
  Getting just the standard model at intersecting branes,
  JHEP {\bf 0111} (2001) 002
  [hep-th/0105155].
  
\bibitem{Blumenhagen:2001te}
  R.~Blumenhagen, B.~Kors, D.~Lust and T.~Ott,
  ``The standard model from stable intersecting brane world orbifolds,''
  Nucl.\ Phys.\ B {\bf 616} (2001) 3
  [arXiv:hep-th/0107138].
  
  
\bibitem{Abel:2003yh}
  S.~A.~Abel, O.~Lebedev and J.~Santiago,
  Flavour in intersecting brane models and bounds on the string scale,
  Nucl.\ Phys.\ B {\bf 696} (2004) 141
  [hep-ph/0312157].




\bibitem{Blumenhagen:2001mb}
  R.~Blumenhagen, B.~K\"ors, D.~L\"ust and T.~Ott,
  Intersecting brane worlds on tori and orbifolds,
  Fortsch.\ Phys.\  {\bf 50} (2002) 843
  [hep-th/0112015].




\bibitem{Cvetic:2001tj}
  M.~Cveti\v{c}, G.~Shiu and A.~M.~Uranga,
  Three-family supersymmetric standard like models from intersecting  brane
  worlds,
  Phys.\ Rev.\ Lett.\  {\bf 87} (2001) 201801
  [hep-th/0107143].




\bibitem{Cvetic:2001nr}
  M.~Cveti\v{c}, G.~Shiu and A.~M.~Uranga,
  Chiral four-dimensional N = 1 supersymmetric type IIA orientifolds from
  intersecting D6-branes,
  Nucl.\ Phys.\ B {\bf 615} (2001) 3
  [hep-th/0107166].



\bibitem{Bailin:2006zf}
  D.~Bailin and A.~Love,
  ``Towards the supersymmetric standard model from intersecting D6-branes on
  the Z'(6) orientifold,''
  Nucl.\ Phys.\  B {\bf 755} (2006) 79
  [arXiv:hep-th/0603172].

\bibitem{Camara:2005dc}
  P.~G.~Camara, A.~Font and L.~E.~Ibanez,
  JHEP {\bf 0509} (2005) 013
  [arXiv:hep-th/0506066].
  
  \bibitem{BLZ612} 
   D.~Bailin and A.~Love, in preparation.
  
  \end{thebibliography}
\end{document}